\documentclass[aps,pra,twocolumn,secnumarabic,floatfix]{revtex4-2}
\usepackage[dvipsnames]{xcolor}
\usepackage{tikz}
\usetikzlibrary{arrows.meta,positioning}

\usepackage{graphicx}			
\usepackage{verbatim}			
\usepackage{mathtools}			
\usepackage[makeroom]{cancel}		
\usepackage{dcolumn}			
\usepackage{hyperref}			
\usepackage{color}			
\usepackage{hyperref}
\usepackage{placeins}

\newcommand{\beq}{\begin{equation}}
\newcommand{\eeq}{\end{equation}}

\begin{document}

\title{The Unreasonable Effectiveness of Physics in Biology}
\author{Alexey Burov}
\email{burov@fnal.gov}
\affiliation{Fermilab, PO Box 500, Batavia, IL 60510-5011, USA}
\author{Alexei Tsvelik}
\affiliation{
Brookhaven National Laboratory, Upton, NY 11973-5000, USA}

\begin{abstract}

We demonstrate that the system of fine-tuning constraints for life is overdetermined: the number of inequalities is too large relative to the number of free parameters in the chemical sector and likely not only there. This entails that life-permitting conditions are secured at the level of physics not only by the ``fine-tuned'' fundamental constants, but also by a very special form of the laws. In 1960, Eugene Wigner published his essay “The Unreasonable Effectiveness of Mathematics in the Natural Sciences,” summarizing his reflections on the miraculous comprehensibility of the mathematical structure of physical laws. The present paper points to another, no less remarkable, aspect of this structure—one that, following Wigner’s pattern, may be called the \it{unreasonable effectiveness of physics in biology.}  

\end{abstract}


\maketitle

 


\section{\label{sec:Int}Introduction}

 Fine-tuning of the universe for life refers to the delicate dependence of the physical possibility of life on the fundamental characteristics of the universe, “notably on the form of the laws of nature, on the values of certain constants of nature, and on aspects of the universe’s conditions in its very early stages” \cite{SEP}. This formulation of the fine-tuning problem has a venerable history, tracing back to early design arguments; see, e.g., \cite{SEP,BarrowTipler} for historical discussion. Among the three fundamental characteristics of the universe listed above—the form of the laws, the values of the fundamental constants, and the initial conditions of the Big Bang—the least attention has so far been devoted to the first, namely the structure of the laws themselves. To the best of our knowledge, this attention has largely been limited to discussions of the necessity of the weak interaction for life, a question that remains under debate \cite{Harnik,Barnes,Grohs}. In this paper, we propose a new perspective on the role of the structure of physical laws in the fine-tuning problem.

Discussions of fine-tuning usually tacitly assume that the underlying structure of the laws admits a possibility of such tuning in principle; that is, the system of constraints under consideration is feasible. 

At first sight, questioning such an assumption may seem strange: after all, our very existence appears to be direct evidence that a solution exists. The question, however, is not whether the fine-tuning constraints are satisfied, but what makes their satisfaction possible. In other words, how mathematically generic—or, conversely, how exceptional—is our structure of the physical laws that ensures the feasibility of the system of life-permitting constraints? Would life-permitting tuning with a fixed number of available “knobs”—the fundamental constants—be impossible unless the very structure of the laws were highly special?

We approach this problem in several steps. First, in Sec.~\ref{sec:Math}, we provide a brief overview of the mathematical theory of feasibility for systems of inequalities, showing that Cover’s classical estimate for the feasibility threshold \cite{Cover} gives a reasonable upper bound for a broad class of such systems. Estimates of this kind, though necessarily rough, allow one to grasp the “scale of the problem,” i.e., the order of magnitude of the probability that a generic system of a large number of inequalities admits a solution.

In the approach adopted here, it is important to distinguish between \textit{a posteriori} and \textit{a priori} feasibility. The \textit{a posteriori} question concerns the existence of values of the tunable fundamental constants that satisfy a given set of life-permitting constraints, once the form of the laws has already been specified. Clearly, for our form of physical laws, such a system of constraints is feasible; that is, it is feasible \textit{a posteriori}. By contrast, the \textit{a priori} question concerns a probabilistic estimate of whether a generic system with the same number of inequalities and the same number of tunable ``constants'' is likely to be feasible. If this probability is low, we shall say that such a system of constraints is \textit{a priori} infeasible. In this paper, we show that a generic system with so many constraints and so few tunable constants—the situation encountered in our Universe for our form of life—is \textit{a priori} infeasible.

In Sec.~\ref{Sec:FTGen}, we apply Cover’s probabilistic formulas to the fine-tuning constraints discussed in the literature in the contexts of particle physics, nuclear physics, and cosmology. Unfortunately, no well-defined list of such inequalities exists at present that would allow their number to be counted with sufficient accuracy. In view of this, we leave such a count for future work and restrict ourselves here to the hypothesis that this number substantially exceeds the Cover threshold. If this is indeed the case, the a priori probability of feasibility of the resulting system of inequalities is extremely low.

After that, in Sec.~\ref{Sec:Chem}, we consider separately the sector of chemical constraints. Focusing on the bonds formed by the biologically most important quartet of atoms—hydrogen, carbon, nitrogen, and oxygen—we find that, for roughly a hundred constraints, only a single free parameter is available. In this situation, it is hardly meaningful to speak of fine-tuning “knobs”: there are next to none.

In Sec.~\ref{Sec:Obj}, we respond to the most interesting critical questions and objections raised by our colleagues and the referees of this paper.  

We conclude this Introduction with a few comments on the title of the paper. The latter follows the pattern of Eugene Wigner’s 1960 essay, \textit{The Unreasonable Effectiveness of Mathematics in the Natural Sciences}, which highlights the astonishing scope and precision of the mathematical comprehensibility of nature. The essay ends on the following ``cheerful note'':
\begin{quote}
    ``The miracle of the appropriateness of the language of mathematics for the formulation of the laws of physics is a wonderful gift which we neither understand nor deserve. We should be grateful for it and hope that it will remain valid in future research and that it will extend, for better or for worse, to our pleasure, even though perhaps also to our bafflement, to wide branches of learning.'' \cite{Wigner}
\end{quote}

Wigner’s comprehensibility of nature rests on the simplicity of the \textit{mathematical principles of natural philosophy}, to invoke the title of Isaac Newton’s great work. This simplicity implies, among other things, a sufficiently small number of dimensionless fundamental constants. On the other hand, the numerous physical conditions required for life, by contrast, call for a sufficiently large number of the fundamental constants, the ``tuning knobs'' in the laws. Thus, a tension arises between two sets of requirements pulling in opposite directions: the physical requirements for life and the requirements for the discoverability of the very physics that makes life possible. For mathematics to display its Wignerian effectiveness in the natural sciences, the form of physical laws must reconcile two opposing requirements: it must be mathematically simple and, at the same time, ``biologically effective.'' If the former property is ``unreasonable,'' then, by the same token, so is the latter. 

In Sec.~\ref{Sec:Conclusion}, we propose an integrated view of the character of the laws of nature, in which the possibility of fine-tuning the constants for life and Wigner’s \textit{unreasonable effectiveness of mathematics in the natural sciences} are brought together within a single framework.

\section{\label{sec:Math}Overdetermined sets of inequalities}

In this section, we briefly review the main results of the Cover model \cite{Cover}, explaining why it can be regarded as an upper bound on the number of the fine-tuning constraints compatible with the feasibility of the system.

Let us consider a set of $N_r$ restrictions as linear homogeneous inequalities in $N_v$ variables $x_i$, with $i=1,2,\dots,N_v$, where $N_r \geq N_v$, and the coefficients $b_{il}$ are drawn independently from a common probability distribution symmetric about zero:
\begin{equation}
\sum_i b_{il} x_i > 0, \qquad l=1,\cdots,N_r .
\end{equation}
The probability that such a system is feasible was studied in the pioneering work of Cover \cite{Cover}, where a universal result for that value was obtained,
\begin{equation}
\label{Eq:CoverP}
P(N_r,N_v)=2^{-(N_r-1)} \sum_{k=0}^{N_v-1} \binom{N_r-1}{k}.
\end{equation}
It follows, in particular, that in the absence of excessive inequalities, when $N_r = N_v$, the system is always feasible, $P(N_v,N_v)=1$. Once excess constraints appear, $N_e \equiv N_r - N_v > 0$, feasibility is no longer guaranteed. However, for a large number of variables, the probability decreases only gradually with $N_e$ until the so-called Cover threshold, $N_e = N_v$, is reached, where $P(2N_v,N_v)=1/2$. Beyond this point, the probability of feasibility rapidly decreases, following a Gaussian law in the vicinity of the threshold; at $N_v \gg 1$,

\begin{equation}
P(2N_v+\Delta N, N_v)
\approx
\Phi \left(-\frac{\Delta N}{\sqrt{2N_v}}\right),
\end{equation}

where $\Phi$ is the cumulative distribution function of the standard normal distribution. 

Let us now generalize further by allowing the right-hand sides of the constraints to be nonzero random variables $a_l$, all drawn from the same even probability distribution. Because Cover’s homogeneous problem is scale-invariant, any of its solutions can be multiplied by a sufficiently large factor so as to become a solution of the corresponding inhomogeneous problem. In other words, the feasibility properties remain the same as in the homogeneous case.

This, in turn, suggests another possible modification of the system: fixing the norm of the vector, $\|\boldsymbol{x}\|$. In that case, one can introduce normalized variables $\boldsymbol{y} = \boldsymbol{x}/\|\boldsymbol{x}\|$, which effectively rescales all right-hand sides by the same factor, $a_l \to a_l/\|\boldsymbol{x}\|$. In the limit $\|\boldsymbol{x}\| \to \infty$, Cover’s homogeneous results are recovered, and the feasibility threshold for $N_r/N_v$ (the so-called capacity) is $\alpha = 2$. In the opposite limit, $\|\boldsymbol{x}\| \to 0$, constraint $l$ becomes trivial if $a_l < 0$, and unsatisfiable if $a_l > 0$. Constraints of the former kind drive the capacity up, whereas those of the latter kind drive it to zero. After discarding the trivial constraints, one is left with some number of unsatisfiable ones. Hence, the net result is $\alpha \to 0$, unless all $a_l$ are negative. In the latter case, when all constraints become trivial, $\alpha \to \infty$.

Problems of this kind have also been studied for homogeneous constraints on the quadratic-form type. Under assumptions on the random matrices analogous to those used in the linear case, it has been found—by means of rather involved numerical calculations—that the corresponding capacity is significantly smaller than the Cover value,
\begin{equation}
\alpha \approx 1.1\text{--}1.3,
\end{equation}
see, e.g., \cite{Thrampoulidis}.

This points to the conjecture that the Cover capacity, $\alpha=2$, serves as a general upper bound for the feasibility threshold of broad classes of systems of inequalities. However, this threshold can be exceeded when the system approaches degeneracy, in the sense that some constraints become nearly redundant, being \textit{a priori} implied by others. In such cases, the effective number of independent constraints is reduced, and the formally defined capacity can, in principle, become arbitrarily large. 

In the context of the problem addressed in this paper, there is no reason to assume such \textit{a priori} degeneracy, let alone strong degeneracy. Each constraint is associated with a distinct physical process, often rather specific to that requirement; hence, there are no grounds for expecting generic redundancy among the requirements. On the contrary, as mentioned above, the capacities of some generic nonlinear systems should be expected to be lower than the Cover threshold.

Thus, we conclude that the Cover model provides a reasonably justified estimate of the a priori feasibility of the system of life-permitting constraints with a given number of tunable parameters.

\section{\label{Sec:FTGen} Fine-Tuning: Everything but Chemistry}

\begin{figure}[htbp]
    \centering
    \includegraphics[width=1.0\linewidth]{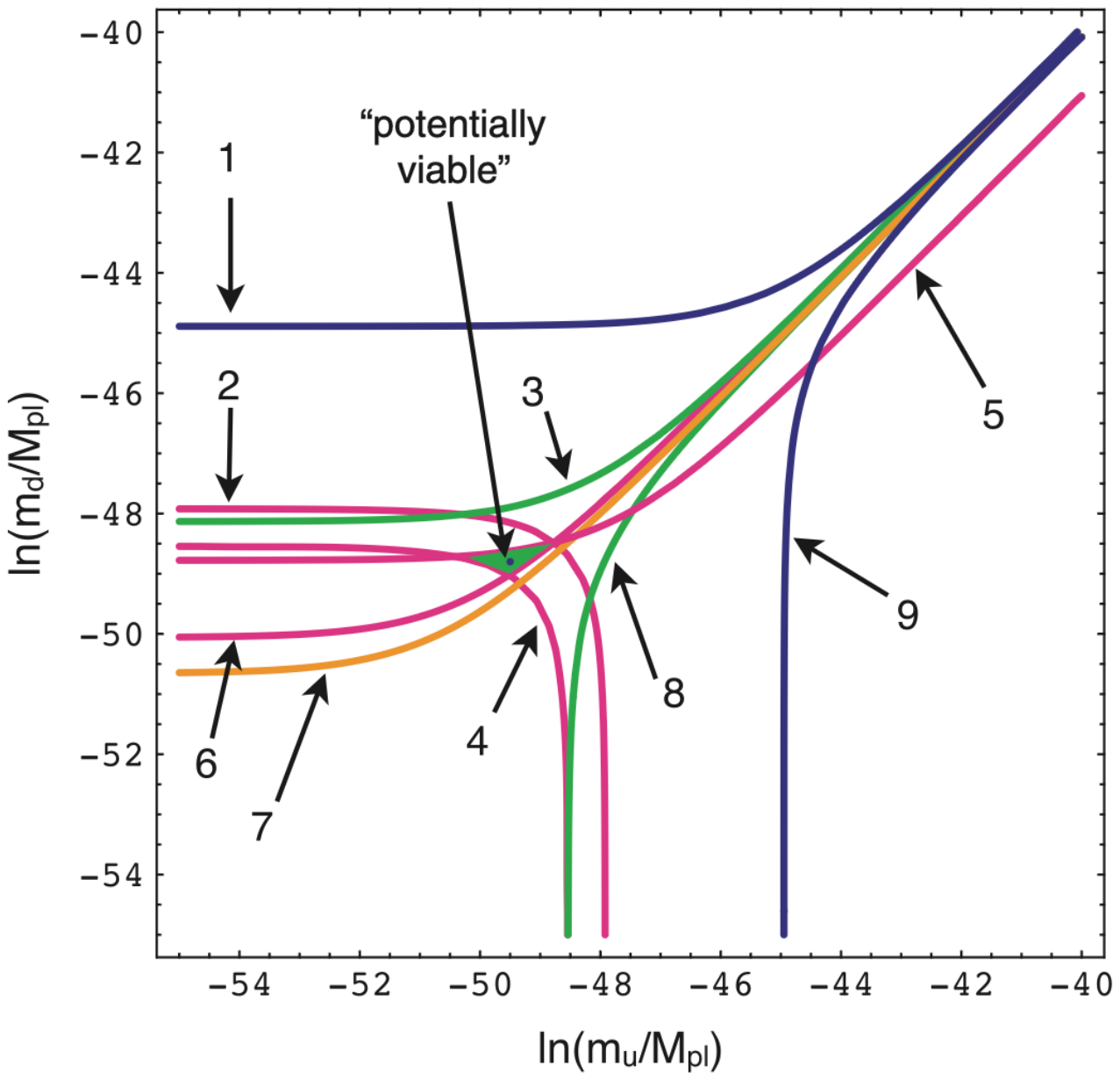}
    \caption{The boundaries of life-permitting constraints for up- and down-quark masses, $m_u$ and $m_d$, in the units of Planck mass $M_{pl}$; computed in Ref.~\cite{Barr}. The arrows (as depicted in Ref.~\cite{Barnes}) indicate the allowed side of each curve. The blue dot inside the potentially viable green area marks the actual values. Figure courtesy of S.~Barr.}
    \label{fig:Barr}
\end{figure}

In this section, we estimate the probability of the joint feasibility of conventional fine-tuning requirements arising from all areas of physics and cosmology, with the exception of chemistry, which will be discussed separately in the next section.

A representative example of apparent overdetermination is shown in Fig.~1, taken from Ref.~\cite{Barnes}. The figure displays nine viability constraints on the masses of the two lightest quarks, which constitute protons and neutrons; the explanations of these constraints can be found in the cited publications. The central green region indicates where all nine constraints are simultaneously satisfied; its smallness suggests that the very existence of a potentially viable domain is fragile. The blue dot inside the green area marks the actual values.

Note that, in addition to the two quark masses, the relevant fundamental parameters underlying the nine curves include four additional key constants of the Standard Model: three interaction couplings and the electron mass. If these four parameters were effectively fixed for some reason, the Cover approach would suggest that the system is most likely overdetermined: for $N_v=2$ and $N_r=9$, Eq.~\ref{Eq:CoverP} gives the feasibility probability $P=0.035$. By contrast, if these four parameters are actually free, then Fig.~\ref{fig:Barr} represents only a two-dimensional slice of a higher-dimensional parameter space with $N_v=6$. In this case, the Cover threshold yields $N_r=12$, and the system is more likely to be feasible than not. 

Let us now turn to a more general analysis of fine-tuning in particle physics and cosmology. To begin with, of the approximately thirty dimensionless fundamental constants, only about a dozen, $N_v = 12$, currently enter the known fine-tuning constraints \cite{Barnes,Adams,Sloan}. 

To the best of our knowledge, as of July 2026 no single generally accepted list of fine-tuning constraints exists. Under these circumstances, we consider it preferable to refrain here from giving a definite estimate of the number of such constraints, leaving this task for future work. Nevertheless, we believe that it may be useful to formulate a hypothesis based on our preliminary estimates of the material collected in the review literature \cite{BarrowTipler,Barnes,Adams,Sloan}: the number of constraints substantially exceeds the Cover threshold $2N_v \approx 24$, thereby rendering the system overdetermined. Moreover, there is every reason to expect the number of known constraints to continue to grow: while previously identified requirements tend to remain in force, new ones are likely to be added.

\section{\label{Sec:Chem} Almost Nothing to Tune: Chemistry}   

Let us first determine how many relevant dimensionless constants—the “knobs” of fine-tuning—are present in the chemical sector of the life-permitting constraints.

The chemical properties of atoms and molecules are governed by the Schrödinger equation and the parameterless Pauli principle. The former is essentially the nonrelativistic wave equation for electrons in the Coulomb fields of one another and of one or more nuclei. 
If one adopts atomic units, in which the proton charge $e$, the electron mass $m_e$, and Planck’s constant $\hbar$ are set equal to unity, then, to leading order in the small relativistic parameter $\alpha_e \equiv e^2/\hbar c$ (where $c$ is the speed of light) and in the small electron–proton and electron–neutron mass ratios, the Schrödinger equation contains no free parameters whatsoever.

Consequently, to leading order in these small parameters, all chemical properties—bond energies and lengths, as well as molecular dipole moments—are fixed up to universal scaling factors. In this sense, dimensionless quantities—such as ratios of bond energies, relative sizes of atoms and molecules together with bond angles and other form factors—do not scale at all; they are fully determined by the mathematical form of the dimensionless nonrelativistic Schrödinger equation and the Pauli principle. As Ref.~\cite{king} puts it, ``The solutions to the scaled electronic Schrödinger equation are pure numbers independent of the fundamental constants.'' It is difficult to imagine a more striking example of an overdetermined system, with a lot of requirements and almost no knobs to tune.
\FloatBarrier
The foregoing requires some qualification: 
\begin{itemize}
\item The possibility of life requires that the energy scale of chemical bonds, set by $m_e c^2 \alpha_e^2$, be compatible with the ambient temperature $T$, giving rise to the dimensionless parameter $\tau = T/(m_e c^2 \alpha_e^2)$.
\item The Schrödinger equation for electrons in the Coulomb fields of one another and of the nuclei neglects relativistic effects. At next order in $\alpha_e^2$, these effects give rise to fine structure—the splitting of Coulomb energy levels due to spin–orbit interaction. For light atoms (H, C, N, O), this splitting is negligible.
\item To leading order, chemical calculations can be carried out in the approximation of infinitely heavy nuclei. However, this approximation requires corrections for some moderately heavy nuclei, where vibrational degrees of freedom may play a role. Such effects introduce corrections governed by the electron-proton mass ratio, $\beta=m_e/m_p \ll 1$. The proton-neutron mass ratio is fixed by nuclear physics to be close to unity with such accuracy that any deviations are insignificant. The requirements that the fine–structure and the mass ratio be small, $\alpha_e, \beta \ll 1$, are imposed by cosmology, nuclear and atomic physics, see Ref.~\cite{Barrow2002Constants}, pp. 167, 168.       
\end{itemize}
Let us estimate now the number of life–permitting constraints in the chemical sector. Modern molecular biology identifies 20 or more elements involved in the vital processes of higher animals \cite{Remick2023}. However, for the purposes of this paper it suffices to restrict attention to the first four—hydrogen, carbon, nitrogen, and oxygen—and to the bonds between them. For the observed values $\alpha_e \approx 1/137$ and $\beta \approx 1/1800$, relativistic effects and finite nuclear mass effects are negligible for these elements; see the article of R.~King et al.~\cite{king} for details. To make these effects significant, the corresponding relativistic and mass-term factors would have to exceed their natural values by one to two orders of magnitude, as demonstrated by the extensive simulations of King et al. This is why they observe that ``the broad $(\alpha_e, \beta)$ sensitivity determined here for the chemistry of life-supporting molecules is not as spectacular as the narrow constraints on these fundamental constants established previously in physics and cosmology.'' Indeed, when life-permitting constraints are, to leading order, fixed by parameterless equations, it hardly can be otherwise.
 
Within the quartet of H, C, N and O, one finds twelve biologically essential covalent bonds \cite{Nelson}:
(N--H, O--H, C--H, C--C, C{=}C, C--N, C{=}N, C$\equiv$N, C--O, C{=}O, N--O, O--O),
six hydrogen bonds \cite{Jeffrey}
(N--H$\cdots$H, O--H$\cdots$H, C--H$\cdots$N, C--H$\cdots$O, N--H$\cdots$N, N--H$\cdots$O),
and $4\cdot5/2=10$ van der Waals interactions, for a total of 28 distinct bonds.

Biological requirements generally impose multiple two-sided constraints on each of these bonds—constraints on their energies, equilibrium lengths, dipole moments, and the angles the bonds form. The satisfaction of these requirements underlies the remarkable ability of these atoms to serve as elements of a kind of Lego construction set for assembling molecular machines of enormous complexity and size. The resulting number of inequalities associated with these bonds appears to be of order one hundred. For the probability of joint feasibility of a system of roughly a hundred inequalities not to be negligibly small, one would require at least several tens of adjustable variables, whereas there is only one here, the overall scaling factor $\tau$.

Note that the elements H, C, N, and O, which constitute more than $99\%$ of the atoms in the human body, belong to four different chemical groups. Therefore, one should not expect the inequalities associated with these bonds to be {\it a priori} dependent; in other words, the system should not be expected to be {\it a priori} degenerate, let alone significantly so.

Thus, consideration of the chemical sector demonstrates that approximately $N_r \simeq 100$ constraints on bonds among H, C, N, and O atoms are met with only a single temperature parameter. Without any tuning of the fundamental constants, essential biochemistry follows from the parameter-free forms of Coulomb’s law, the Schrödinger equation, and the Pauli principle—as Athena sprang fully armed from the head of Zeus.

\section{\label{Sec:Obj} Questions and Objections}

In discussions with colleagues and referees concerning the ideas behind this work and earlier drafts of the manuscript, we have received a number of questions and critical comments on the reasoning developed here; the most interesting of these are addressed in this section.

\begin{enumerate}
    \item Are the constraints dependent?
    \begin{itemize}
        \item \textbf{Objection.} The properties of covalent, hydrogen-bond, and van der Waals interactions among the four elements discussed by the authors are not, in general, independent of one another. As a result, the number of independent constraints may be significantly smaller than the order one hundred estimate suggested by the authors.
        
        \item \textbf{Response.} Since the parameters of the bonds in any case are computed from one and the same set of laws, these parameters are generically correlated, or interdependent. But this interdependence of bond parameters by no means implies that the biochemical matching requirements on the parameters of this “Lego-like” construction set will be generically satisfied. The law that generates the elements of this construction set, together with their bonds, is one thing; the biochemical life-permitting requirements imposed on these bonds are quite another. These constraints are external to the logic of law and are not generically secured by it. 
        
        Let us also stress once again that a priori and a posteriori feasibility should not be conflated. For the existing form of the laws, the constraints under discussion are known to be satisfied, so the corresponding system of constraints is feasible \textit{a posteriori}. From the \textit{a priori} perspective, however, the possibility of satisfying so many constraints for generic laws with only one free parameter would be extraordinarily unlikely—or, in Wigner’s usage, \textit{unreasonable}.
    \end{itemize}
    
    \item Is the combinatorial factor important?
    \begin{itemize}
        \item \textbf{Objection.} The requirement for viable organic chemistry is not that these four particular elements, with charges 1, 6, 7, and 8, must have suitable properties, but rather the much weaker requirement that, among roughly 100 stable elements, there exists at least one four-element subset with suitable properties. Since the number of four-element subsets of a 100-element set is of order $10^6$, this is an important distinction. 
        
        \item \textbf{Response.} In principle, this objection may be accepted, setting aside the complications associated with heavier biochemistry and the relative rarity of heavier elements. However, a factor of a million does not affect the conclusion, since the Cover probability is far too small: for, say, one hundred inequalities with one free parameter, $P(100,1) \simeq 10^{-30}$.    
    \end{itemize}

    \item What is the role of evolutionary adaptation?
    \begin{itemize}
        \item \textbf{Objection.} Perhaps biological evolution can adapt the shape and composition of living beings to the properties of chemical interactions, thereby effectively increasing the probability of feasibility. How could evolutionary factors affect the conclusions of this paper?
        
        \item \textbf{Response.} Evolution can exploit whatever chemically viable structures are available and thereby relax some requirements. This freedom presupposes a sufficiently rich chemical basis: stable yet cleavable bonds, molecular diversity, long information-bearing molecules, catalysis, energy conversion, and so on. Evolution can optimize within such a landscape, but it cannot create the landscape itself. Perhaps evolutionary adaptability may reduce the effective number of independent constraints. For the same number of the free fundamental parameters, it would affect the main conclusion only if it reduced the effective number of constraints by more than an order of magnitude. In view of the extraordinary complexity of living organisms, we see no reason to expect such a dramatic reduction in the number of life-permitting constraints as a generic possibility. With a substantially larger number of free fundamental constants, many forms of the laws could, in principle, satisfy the requirements for life. But such laws would be mathematically much more complex and therefore unlikely to be discoverable by emerging sapient beings. In other words, it would be quite unlikely that mathematics would be \textit{effective} in Wigner’s sense in such worlds. Even our “simple” laws, judging by the history of their discovery, were hard-won at the very limit of the intellectual capacities of the great founders of mathematical physics.
    \end{itemize}

\end{enumerate}

\section{\label{Sec:Conclusion} Conclusion}

The known laws of physics exhibit two fundamental and widely discussed features. The first is their mathematical simplicity and structural elegance, underlying Eugene Wigner’s celebrated ``unreasonable effectiveness of mathematics in the natural sciences'' \cite{Wigner}, i.e., the discoverability of the Universe across an enormous range of parameters and with extraordinary precision; see also further development of this issue by Mark Steiner \cite{Steiner}. The second, no less striking feature is the fine-tuning of the constants of these laws for life, including intelligent life, see e.g. \cite{BarrowTipler,Rees2000JustSixNumbers,Barrow2002Constants,BarnesLewis2016FortunateUniverse,Adams,Sloan}. 

The structure of the laws, understood as the mathematical principles of physics, has traditionally been associated with their discoverability, whereas the theoretically free, empirically determined constants of the laws have been considered in the context of enabling life; symbolically, {\it structure $\Rightarrow$ discoverability, constants $\Rightarrow$ viability}. 

In addition to these two relations, Robin Collins has proposed a crosswise hypothesis, namely that the fundamental constants are fine-tuned not only for life, but also for the discoverability of the laws themselves \cite{Collins2018Discoverability}; symbolically, {\it constants $\Rightarrow$ discoverability}.

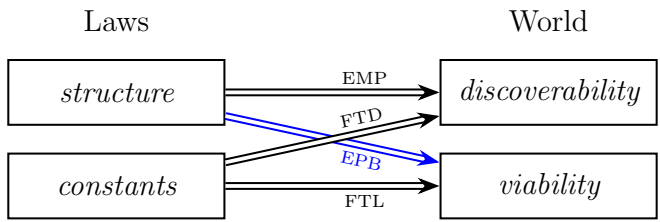
\begin{figure}[htbp]
\centering
\resizebox{\columnwidth}{!}{%
\begin{tikzpicture}[
>=Stealth,
box/.style={
draw,
thick,
rectangle,
minimum width=3cm,
minimum height=0.9cm,
align=center,
font=\large\itshape
}
]

\node[font=\large] at (0,2.0) {Laws};
\node[font=\large] at (6,2.0) {World};

\node[box] (S) at (0,1) {structure};
\node[box] (C) at (0,-0.3) {constants};

\node[box] (D) at (6,1) {discoverability};
\node[box] (V) at (6,-0.3) {viability};

\draw[->,thick,double,double distance=1.5pt]
(S) -- node[pos=0.65,above,font=\scriptsize] {EMP} (D);

\draw[->,thick,double,double distance=1.5pt]
(C) -- node[pos=0.65,below,font=\scriptsize] {FTL} (V);

\draw[->,thick,double,double distance=1.5pt,color=blue]
(S) -- node[pos=0.65,sloped,below,font=\scriptsize,color=blue] {EPB} (V);

\draw[->,thick,double,double distance=1.5pt]
(C) -- node[pos=0.65,sloped,above,font=\scriptsize] {FTD} (D);

\end{tikzpicture}%
}
\caption{Relations between the properties of the laws and the capacities of the world. EMP stands for the Effectiveness of Mathematics in Physics, EPB for the Effectiveness of Physics in Biology (blue arrow, proposed in this work), FTL for Fine-Tuning for Life, and FTD for Fine-Tuning for Discoverability (proposed by R.~Collins \cite{Collins2018Discoverability}).}
\label{Fig:LawsWorld}
\end{figure}

In this paper, we aim to demonstrate the second, hitherto unnoticed cross-link: \textit{structure $\Rightarrow$ viability}. The full set of links is shown schematically in Fig.~\ref{Fig:LawsWorld}. Namely, we argue that the very possibility of fine-tuning the constants within a mathematically simple and elegant structure of the laws appears, in a purely mathematical sense, highly implausible, owing to the tension between the large number of constraints and the disproportionately small number of tunable free parameters available to satisfy them.

This idea was put forward by one of us more than a decade ago, giving rise to the title of his book \textit{Life in the Impossible World} \cite{TsvelikImpossible2012}; here we attempt to formulate that idea as a quantitative argument.

The principle unifying all the above features of the physical laws may be expressed as a \textit{minimax of complexity} required for their discoverability \cite{GPT}. Were the laws simpler, life-supporting fine-tuning would be very unlikely—indeed, even at the present level of complexity, such fine-tuning appears to be a mathematically unsolvable problem, as we tried to demonstrate in this paper. Were they more complex, we, emerging sapient beings, would likely be unable to discover them—already at the current level, the laws seem to have been discovered at the very limits of the capabilities of the most gifted pioneers.

In the present paper, we have attempted to show that the physical laws are even more remarkable than is usually acknowledged. Are other worlds possible, with different laws, in which mathematics would possess Wignerian effectiveness, and in which physics would therefore be not only mathematically elegant but also biologically effective? We have tried to demonstrate that the tension between these requirements is so strong that the mathematical effectiveness of our world seems even more “unreasonable” than Eugene Wigner apparently thought.

\begin{acknowledgments}
We are grateful to our long-standing interlocutors Mikhail Arkadev and Lev Burov for many inspiring discussions and constructive criticism. We are indebted to Robin Collins for his valuable comments, and especially thankful to both of our referees for their numerous suggestions and, most of all, for their challenging questions.
\end{acknowledgments}

\bibliography{bibfile}			

@misc{SEP,
  title = {Fine-Tuning},
  author = {S. Friederich},
  howpublished = {\emph{Stanford Encyclopedia of Philosophy}},
  note = {\url{https://plato.stanford.edu/entries/fine-tuning/}},
  year = {2020}
}

@book{BarrowTipler,
  editor = {John D. Barrow and Frank J. Tipler},
  title = {The Anthropic Cosmological Principle},
  publisher = {Oxford  University Press},
  address = {New York},
  year = {1986}
}

@article{Barnes,
  author = {L. A. Barnes},
  title = {The Fine-Tuning of the Universe for Intelligent Life},
  journal = {Publ. Astron. Soc. Aust.},
  volume = {29},
  pages = {529--564},
  year = {2012},
  doi = {10.1071/AS12015}
}

@article{Harnik,
  author = {R. Harnik and G. D. Kribs and G. Perez},
  title = {A universe without weak interactions},
  journal = {Phys. Rev. D},
  volume = {74},
  pages = {035006},
  year = {2006},
  doi = {10.1103/PhysRevD.74.035006}
}

@article{Grohs,
  author = {E. Grohs and A. R. Howe and F. C. Adams},
  title = {Universes without the weak force: Astrophysical processes with stable neutrons},
  journal = {Phys. Rev. D},
  volume = {97},
  pages = {043003},
  year = {2018},
  doi = {10.1103/PhysRevD.97.043003}
}

@book{Sloan,
  editor = {D. Sloan and R. A. Batista and M. T. Hicks and R. Davies},
  title = {Fine-Tuning in the Physical Universe},
  publisher = {Cambridge University Press},
  address = {Cambridge},
  year = {2020}
}

@article{Barr,
  author = {S. M. Barr and A. Khan},
  title = {Anthropic tuning of the weak scale and of {$m_u/m_d$} in two-Higgs-doublet models},
  journal = {Phys. Rev. D},
  volume = {75},
  pages = {045002},
  year = {2007},
  doi = {10.1103/PhysRevD.75.045002}
}

@article{Adams,
  author = {F. C. Adams},
  title = {The degree of fine-tuning in our universe---and others},
  journal = {Phys. Rep.},
  volume = {807},
  pages = {1--111},
  year = {2019},
  doi = {10.1016/j.physrep.2019.02.001}
}

@article{Cover,
  author = {T. M. Cover},
  title = {Geometrical and Statistical Properties of Systems of Linear Inequalities with Applications in Pattern Recognition},
  journal = {IEEE Trans. Electron. Comput.},
  volume = {EC-14},
  number = {3},
  pages = {326--334},
  year = {1965},
  doi = {10.1109/PGEC.1965.264137}
}

@article{Thrampoulidis,
  author = {C. Thrampoulidis and S. Oymak and B. Hassibi},
  title = {The Gaussian Min--Max Theorem in the Context of Convex Optimization},
  journal = {Found. Trends Signal Process.},
  volume = {12},
  number = {1},
  pages = {1--159},
  year = {2018},
  doi = {10.1561/2000000083}
}

@article{Wigner,
  author = {E. Wigner},
  title = {The Unreasonable Effectiveness of Mathematics in the Natural Sciences},
  journal = {Commun. Pure Appl. Math.},
  volume = {13},
  number = {1},
  pages = {1--14},
  year = {1960},
  doi = {10.1002/cpa.3160130102}
}

@incollection{GPT,
  author    = {A. Burov and L. Burov},
  title     = {Genesis of a Pythagorean Universe},
  booktitle = {Trick or Truth? The Mysterious Connection Between Physics and Mathematics},
  editor    = {A. Aguirre and B. Foster and Z. Merali},
  publisher = {Springer},
  address   = {Switzerland},
  pages     = {157--171},
  year      = {2016},
  doi       = {10.1007/978-3-319-27495-9},
  eprint        = {1411.7304},
  archivePrefix = {arXiv},
  primaryClass  = {physics.hist-ph},
  url={https://arxiv.org/abs/1411.7304}
}

@article{king,
  author = {R.A.King and A. Siddiqi and W. D. Allen and H. F. Schaefer, III},
  title = {Chemistry  as a Function of the Fine-Structure Constant and the Electron-Proton Mass Ratio},
  journal = {Phys. Rev. A},
  volume = {81},
  pages = {042523},
  year = {2010},
  doi = {10.1103/PhysRevA.81.042523}
}

@book{Nelson,
  author    = {Nelson, David L. and Cox, Michael M.},
  title     = {Lehninger Principles of Biochemistry},
  edition   = {8},
  year      = {2021},
  publisher = {W. H. Freeman},
  address   = {New York}
}

@book{Jeffrey,
  author    = {Jeffrey, George A.},
  title     = {An Introduction to Hydrogen Bonding},
  year      = {1997},
  publisher = {Oxford University Press},
  address   = {New York}
}

@article{Remick2023,
  author  = {Remick, K. and D Helman, J. D.},
  title   = {The Elements of Life: A Biocentric Tour of the Periodic Table},
  journal = {Adv. Microb. Physiol. },
  volume  = {82},
  pages   = {1-127},
  year    = {2023},
  doi     = {10.1016/bs.ampbs.2022.11.001}
}

@incollection{Collins2018Discoverability,
  author    = {Collins, Robin},
  title     = {The Argument from Physical Constants: The Fine-Tuning for Discoverability},
  booktitle = {Two Dozen (or So) Arguments for God: The Plantinga Project},
  editor    = {Walls, Jerry L. and Dougherty, Trent},
  year      = {2018},
  publisher = {Oxford University Press},
  address   = {Oxford},
  pages     = {89--107},
  doi       = {10.1093/oso/9780190842215.003.0006}
}

@book{TsvelikImpossible2012,
  author    = {Tsvelik, Alexei.},
  title     = {Life in the Impossible World (in Russian)},
  year      = {2012},
  publisher = {Ivan Limbakh Publishing Company},
  address   = {SPb, Russia}
}

@book{Barrow2002Constants,
  author    = {Barrow, John D.},
  title     = {The Constants of Nature: From Alpha to Omega---The Numbers That Encode the Deepest Secrets of the Universe},
  publisher = {Pantheon Books},
  address   = {New York},
  year      = {2002}
}

@book{Steiner,
  author    = {Steiner, Mark},
  title     = {The Applicability of Mathematics as a Philosophical Problem},
  year      = {1998},
  publisher = {Harvard University Press},
  address   = {London}
}

@book{Rees2000JustSixNumbers,
  author    = {Rees, Martin J.},
  title     = {Just Six Numbers: The Deep Forces That Shape the Universe},
  publisher = {Basic Books},
  address   = {New York},
  year      = {2000}
}

@book{BarnesLewis2016FortunateUniverse,
  author    = {Barnes, Luke A. and Lewis, Geraint F.},
  title     = {A Fortunate Universe: Life in a Finely Tuned Cosmos},
  publisher = {Cambridge University Press},
  address   = {Cambridge},
  year      = {2016}
}
\end{document}